\documentstyle[prl,aps,12pt]{revtex}

\newcommand{\AmS}{{\protect\the\textfont2

  A\kern-.1667em\lower.5ex\hbox{M}\kern-.125emS}}

\begin{document}
\title{Tree Amplitudes and Linearized SUSY Invariants in D=11 Supergravity}

\vskip 3cm

\author{Domenico Seminara
\thanks{CEE Post-Doctoral Fellow under contract FMRX-CT96-0045}}
\address{Laboratoire de Physique
        Th{\'e}orique de L' {\'E}cole Normale Sup{\'e}rieure
\thanks{Unit{\'e} Mixte associee au Centre de la Recherche Scientifique et
 {\`a} l' {\'E}cole Normale Sup{\'e}rieure},
 24 Rue Lhomond, F-75231, Paris CEDEX 05, France.}
\maketitle
\vskip 2cm
\begin{abstract}
We exploit the tree level bosonic 4-particle scattering amplitudes 
in $D=11$ supergravity~to~construct the bosonic part of a linearized
$\rm supersymmetry-,$ coordinate- and gauge-invariant. By differentiation,
this invariant  can be promoted to be the natural lowest (two-loop) order
counterterm. Its existence implies that the perturbative supersymmetry does
not protect this ultimate supergravity from infinities, given also the
recently  demonstrated divergence of its 4-graviton amplitude.
\end{abstract}
\vfill 
\begin{flushright}
LPTENS 99/60
\end{flushright}
\newpage

After the so-called ``second string revolution'',  D=11 supergravity 
\cite{CrJuSc} has regained its well-deserved central role, being 
a very efficient tool to investigate the (mostly unknown) $M-$theory. 
In particular, it has been realized how computing  loop effects
in  this last supergravity \cite{BDetal}  can shed some light on some 
(protected) sectors of the $M-$theory effective action\cite{Green}. 
These connections add futher motivation to our quest for higher 
order on shell SUSY invariants in $D=11$, whose construction is 
technically difficult to handle, a tensor calculus being absent.

In this brief review, we will supply (the linearized part
of) one such invariant. Our original interest in constructing SUSY
was triggered by a more modest, but intriguing question. We 
wanted to determine unambiguously whether there exist local 
invariants that can serve as counterterms in loop calculations, 
at lowest relevant order. This nontrivial exercise has its roots
in lower-dimensional SUGRAs, where the existence of invariants is
easier to decide. There no miracle seemed to protect this class of  
theories from diverging, since one is always able to single out
a candidate counterterm \cite{desergermany}. However, given 
all the properties unique to
D=11, and the fact that it is the last frontier -- a local QFT
that is non-ghost ({\it i.e.}, has no quadratic curvature terms)
and reduces to GR -- it is sufficiently important not to give up
hope too quickly before abandoning D=11 SUGRA and with it all QFTs
incorporating GR quickly on non-renormalizability grounds. 

The underlying idea is a simple one: the tree level scattering
amplitudes constructed within a perturbative expansion of the
action are {\it ipso facto} globally SUSY and linearized gauge
invariant.  Furthermore, because linearized SUSY means precisely
that it does not mix different powers of fields, the 4-point
amplitudes taken together form an invariant. Finally, the lowest
order bosonic 4-point amplitudes are independent of fermions:
virtual fermions never appear at tree level. 
In order to use
this invariant for counterterm purposes, it will first be
necessary to remove from it the nonlocality  associated with
exchange of virtual graviton and form particles .
Indeed, the task here will be not only to remove nonlocality but
to add sufficient powers of momentum to provide an on-shell
invariant of correct dimension to make it an acceptable 2-loop
counterterm candidate, this (rather than 1 loop) being the first
possible order (by dimensions) where 4 point amplitudes can
contribute. In this way, we will make contact with the conclusive
2-loop results of \cite{BDetal}, where it was possible to exhibit
the infinity of the 4-graviton component of the invariant.
Details  of our result can be in \cite{sdds}.

The basis for our computations is the  full action of
\cite{CrJuSc}, expanded to the order required for obtaining the
four-point scattering amplitudes among its two bosons, namely the
graviton and the three-form potential $A_{\mu \nu\alpha}$ with
field strength $F_{\mu \nu\alpha\beta}\equiv 4\partial_{[ \mu }
A_{\nu\alpha\beta]}$, invariant under the gauge transformations
$\delta A_{\mu \nu\alpha}=\partial_{[\mu }\xi_{\nu\alpha ]}$. From
the bosonic truncation of this action (omitting obvious summation
indices),
\begin{equation}
\label{Lagra} I^B_{11}=\int d^{11}x \left
[-\frac{\sqrt{g}}{4\kappa^2} R(g) -\frac{\sqrt{g}}{48} F^2 \right.
\end{equation}
$$ \left. +\frac{2\kappa}{144^2}\epsilon^{1\cdots 11}
F_{1\cdots}F_{5\cdots}A_{..11}\right], $$
 we extract the relevant
vertices and propagators; note that $\kappa^2$  has dimension
$[L]^9$ and that the (P, T) conserving cubic Chern-Simons (CS)
term depends explicitly on $\kappa$ but is (of course)
gravity-independent. The propagators come from the quadratic terms
in $\kappa h_{\mu \nu}\equiv g_{\mu \nu}-\eta_{\mu \nu}$ and
$A_{\mu \nu\alpha}$; they need no introduction. There are three
cubic vertices, namely graviton, pure form and mixed form-graviton
that we schematically represent as
\begin{eqnarray}
\label{gggl}
&&\!\!\!\!\!\!\!\!\!\!\!\!
V^g_3\simeq(\partial
h\partial h)h\equiv \kappa T^{\mu \nu}_g h_{\mu \nu},\ \
T^{\mu \nu}_g\equiv G^{\mu \nu}_{(2)}(h), \nonumber\\
&&\!\!\!\!\!\!\!\!\!\!\!\!
V_3^{gFF}\equiv
\kappa T^{\mu \nu}_F h_{\mu \nu},\   T^{\mu \nu}_F\equiv F^\mu 
F^\nu-\frac{1}{8}\eta^{\mu \nu} F^2, \\
&&\!\!\!\!\!\!\!\!\!\!\!\!
V_3^F\equiv\kappa
A_{\mu \nu\alpha} C_F^{\mu \nu\alpha}, \
C_F^{\rho\sigma\tau}\equiv
\epsilon^{\rho\sigma\tau\mu _1\dots\mu _8}
F_{\mu _1\dots}F_{\dots\mu _8}.\nonumber
\end{eqnarray}
 The form's current $C_F$ and stress
tensor $T_F$ are both manifestly gauge invariant. In our
computation, two legs of the three-graviton vertex are always on
linearized Einstein shell; we have exploited this fact in writing
it in the simplified form (\ref{gggl}), the subscript on the
Einstein tensor denoting its quadratic part in $h$. 
To achieve coordinate invariance to correct,
quadratic, order one must also include the four-point contact
vertices
$$
V^{g}_4\sim \kappa^2 (\partial h\partial h) h h, \ \ V_4^{gF}
=\kappa^2  \frac{\delta I^F}{\delta g_{\alpha\beta}\delta
g_{\mu \nu}} h_{\alpha\beta} h_{\mu \nu}
$$
when calculating the amplitudes; these are the remedies for the
unavoidable coordinate variance of the gravitational stress tensor
$T^{\mu \nu}_g$ and the fact that  $T^{\mu \nu}_F h_{\mu \nu}$ is
only first order coordinate-invariant. The gravitational vertices
are not given explicitly, as they are both horrible and well-known
\cite{Sannah}. 

We start with the $4-$graviton amplitude, obtained by contracting
two $V^g_3$ vertices in all three  channels (labelled by the
Mandelstam variables $(s,t,u)$) through an intermediate  graviton
propagator (that provides a single denominator);  adding  the
contact $V^g_4$ and then setting the external graviton
polarization tensors on free Einstein shell. The resulting
amplitude $M^g_4 (h)$ will be a nonlocal (precisely thanks to the
local $V^g_4$ contribution!) quartic in the Weyl tensor\footnote{
We do not differentiate in notation between Weyl and Riemann here
and  also express amplitudes in covariant terms for simplicity,
even though  they are only valid to   lowest relevant order in the
linearized curvatures.}. 
Within  our space limitations,
we cannot exhibit  the actual calculation here; fortunately, this
amplitude  has already  been given  (for  arbitrary $D$) in the
pure gravity context \cite{Sannah}. It can be shown, using the
basis of \cite{fulling}, to be of the form
\begin{equation}
\label{pippo}
M^{g}_4=\kappa^2 (4 s t u)^{-1}
t_8^{\mu _1\cdots\mu _8} t_8^{\nu_1\cdots\nu_8}\times 
\end{equation}
$$
R_{\mu _1\mu _2\nu_1\nu_2}R_{\mu _3\mu _4\nu_3\nu_4}R_{\mu _5\mu _6\nu_5\nu_6}
R_{\mu _7\mu _8\nu_7\nu_8}\equiv   \frac{L^g_4}{s t u},
$$
up to a possible contribution from the quartic Euler density
$E_8$, which is  a total divergence to this order (if present, it
would only contribute at $R^5$ level). The result (\ref{pippo}) is
also the familiar superstring zero-slope limit correction to D=10
supergravity, where the $t_8^{\mu _1\cdots\mu _8}$ symbol originates
from the D=8 transverse subspace\cite{Schwartz}. [Indeed, the
``true'' origin of the ten dimensional analog of (\ref{pippo}) was
actually traced back to D=11 in the one-loop computation of
\cite{Green}.] Note that the local part, $L^g_4$, is simply
extracted through multiplication of $M^g_4$ by $s t u $, which in
no way alters SUSY invariance, because all parts of $M_4$ behave
the same way.

In many respects, the form (\ref{pippo}) for the $4-$graviton
contribution is a perfectly physical one. However in terms of the
rest of the invariant to be obtained below, one would like a
natural formulation with currents that encompass both gravity and
matter in a unified way as in fact occurs in e.g. $N=2$, D=4
supergravity \cite{DK}. This might also lead to some understanding
of other SUSY multiplets. Using the quartic basis expansion, one
may  rewrite  $L^g_4$ in various ways involving conserved $BR$
currents and a closed $4-$form $P_{\alpha\beta\mu \nu}= 1/4
R^{ab}_{[\mu \nu} R_{\alpha\beta]a b}$, for example
$$
L^g_4=48 \kappa^2\left[ 2
B_{\mu \nu\alpha\beta}B^{\mu \alpha\nu\beta}
-B_{\mu \nu\alpha\beta}B^{\mu \nu\alpha\beta}+ \right.
$$
\begin{equation}
P_{\mu \nu\alpha\beta}P^{\mu \nu\alpha\beta}+
\left. 6 B_{\mu \rho\alpha}^{\
\ \ \ \rho}B^{\mu \sigma\alpha}_{\ \ \ \ \sigma} -\frac{15}{49}
(B^{\mu \nu}_{\ \ ~\mu \nu})^2\right]
\end{equation}
with
$B_{\mu \nu\alpha\beta}\equiv R_{(\underline\mu \rho\alpha\sigma}
R^{\ ~\rho\ ~\sigma}_{\underline{\nu})\ ~\beta\ }
-\frac{1}{2}g_{\mu \nu} R_{\alpha\rho\sigma\tau} R_{\beta}^{\
~\rho\sigma\tau}-
\frac{1}{2}g_{\alpha\beta} R_{\mu \rho\sigma\tau}
R_{\nu}^{\ ~\rho\sigma\tau} +\frac{1}{8} g_{\mu \nu}
g_{\alpha\beta}
R_{\lambda\rho\sigma\tau}R^{\lambda\rho\sigma\tau},
$\\
where $(~)$ means symmetrization with weight one of the underlined
indices. 

Let us now turn to the pure form amplitude, whose operative
currents are the Chern-Simons $C^F_{\mu \nu\alpha}$ and the stress
tensor $T^F_{\mu \nu}$, mediated respectively by the $A$ and
graviton propagators; each contribution is separately invariant.
We computed the two
relevant, $C_F C_F$ and $T_F T_F$, diagrams directly, resulting in
the four-point amplitude (see also \cite{12}); 
$M^F_4= (s t u)^{-1} L^F_4= (s t u)^{-1} \kappa^2
(\partial F)^4$, again with an overall ($stu$) factor. An
economical way to organize $L^F_4$ is in terms of matter BR
tensors and corresponding $C^F$ extensions, prototypes being  the
``double gradients'' of $T^F_{\mu \nu}$ and of $C^F$,\\
$
B^F_{\mu \nu\alpha\beta}=
\partial_{\alpha} F_\mu 
\partial_{\beta} F_{\nu}+
\partial_{\beta} F_\mu 
\partial_{\alpha} F_{\nu}-
\frac{1}{4}\eta_{\mu \nu} \partial_{\alpha} F
\partial_{\beta} F,
$\\
$
C^F_{\rho\sigma\tau;\alpha\beta}
=\frac{1}{(24)^2}\epsilon_{\rho\sigma\tau\mu _1\cdots\mu _8}
\partial_\alpha F^{\mu _1\cdots\mu _4} \partial_\beta
F^{\mu _5\cdots\mu _8},
$\\
where $
\partial^\mu B^F_{\mu \nu\alpha\beta}=0,
\ \ \ \partial^\rho
C^F_{\rho\sigma\tau;\alpha\beta}=0.
$.
From the above equation we can construct $L^F_4$ as
$$\!\!\!\!\!
L^F_4=\frac{\kappa^2}{36}B^F_{\mu \nu\alpha\beta}
B^{F}_{\mu _1\nu_1\alpha_1\beta_1}
G^{\mu \mu _1;\nu_1\nu}K^{\alpha\alpha_1;\beta_1\beta}-
$$
\begin{equation}
\label{BF1} \ \
\frac{\kappa^2}{12}
C^F_{\mu \nu\rho;\alpha\beta}C^{F\mu \nu\rho}_{~~~~~~\alpha_1\beta_1}
K^{\alpha\alpha_1;\beta_1\beta}.
\end{equation}
The matrix $G^{\mu \nu;\alpha\beta}\equiv \eta^{\mu \alpha}
\eta^{\nu\beta}+\eta^{\nu\alpha} \eta^{\mu \beta}-
2/9\eta^{\mu \nu} \eta^{\alpha\beta} $ is the usual numerator
of the graviton propagator on conserved sources. The origin of
$K^{\mu \nu;\alpha\beta}\equiv \eta^{\mu \alpha} \eta^{\nu\beta}+
\eta^{\nu\alpha} \eta^{\mu \beta}- \eta^{\mu \nu} \eta^{\alpha\beta}
$ can be traced back to ``spreading" the $stu$ derivatives: for
example, in the $s-$channel, {\it e.g.}, we can write $t u=-1/2
K^{\mu \nu;\alpha\beta}p^1_\mu p^2_\nu p^3_\alpha p^4_\beta$; the
analogous identities for the other channels can be obtained by
crossing\footnote{It is convenient to define $s\equiv(p_1\cdot 
p_2), ~t\equiv(p_1\cdot p_3),~u\equiv(p_1\cdot p_4)$, with
$p_1+p_2=p_3+p_4$.}. It is
these identities that enabled us to write $M_4$'s universally as
$(stu)^{-1} L_4$'s: Originally the $M_4$ have a single denominator
(from the intermediate specific exchange, $s-$, $t-$ or
$u-$channel); we uniformize  them all to $(stu)^{-1}$ through
multiplication of say $s^{-1}$ by $(tu)^{-1} (tu)$. The extra
derivatives thereby distributed  in the numerators have the
further virtue of turning all polarization tensors into curvatures
and derivatives  of forms, as we have indicated. 

The remaining amplitudes are the form  ``bremsstrahlung''
$M^{FFFg}$ and the graviton-form scattering $M^{Fg}_4$.  The
$M^{FFFg}$ amplitude represents radiation of a graviton from one
of the CS  arms, {\it i.e.}, contraction of the  $CS$ and
$T^F_{\mu \nu}h^{\mu \nu}$ vertices by an intermediate $A-$line,
yielding 
\begin{equation}
L_4^{FFF g}=
-\frac{\kappa^2}{3} C^F_{\mu \nu\rho;\alpha\beta}
C^{RF\mu \nu\rho}_{~~~~~~~\alpha_1\beta_1}K^{\alpha\alpha_1;\beta_1\beta},
\end{equation}
where the above current $C^{RF}_{\mu \nu\rho;\alpha\beta}$ is given by\\ 
$4\partial_\lambda\left(
R^{\sigma\  [\lambda }_{\ (\alpha\ ~\beta)} F_{\sigma}^{\
\mu \nu\rho]}\right )-
\frac{2}{3}R^{~\sigma\  ~\lambda }_{\
~(\alpha\ ~\beta)}\partial_\lambda F_{\sigma}^{\ \mu \nu\rho} \; .
$\\
The off-diagonal current $C^{RF}$ has antecedents in $N=2$ D=4
theory \cite{DK}; it is unique only up to terms vanishing on
contraction with $C^F$. 
The $M^{Fg}, \sim\kappa^2  R^2 (\partial F)^2$, has
three distinct diagrams: mixed $T^F T^g$ mediated by the graviton;
gravitational Compton amplitudes $\sim(hh)T_F T_F$ with a virtual
$A-$line, and finally the $4-$point contact vertex $F F h h$. The
resulting  $M_4^{Fg}$ is equal to $(s t u)^{-1} L_4^{Fg}$, where
\begin{eqnarray}
L_4^{Fg}&\!\!\!=\!\!\!&\frac{\kappa^2}{3}
\left (\textstyle{\frac{1}{4}}B^g_{\mu \nu\alpha\beta}
B^F_{\mu _1\nu_1,\alpha_1\beta_1}G^{\mu \mu _1;\nu\nu_1}\right.-\nonumber\\
&&\left. C^{RF}_{\mu \nu\rho;\alpha\beta}
C^{RF\mu \nu\rho}_{~~~~~~\alpha_1\beta_1}\right)
K^{\alpha\alpha_1;\beta_1\beta}, 
\end{eqnarray}
up to subleading terms involving traces. The complete bosonic
invariant,
$
L_{4}  \equiv L_4^F+L_4^g+L_4^{Fg}+L_4^{FFFg},
$
is not necessarily in its  most unified form, but 
we hope to return to this point elsewhere. 

Finally we discuss the
consequences of the very existence of this invariant, for the 
renormalizability properties of $D=11$ supergravity.
With our space limitation, rewiewing the general structure
of the loop expansion and its possible divergences in $D=11$
is an impossible task. For this reason, we will limit ourselves
to a very brief discussion of the one-loop case and we 
go directly to the $2-$loop analysis. For clarity, we choose 
to work in the framework of dimensional  regularization, 
in which only logarithmic divergences appear and consequently
the local counterterm must have dimension zero.\\
\indent
A generic gravitational loop expansion proceeds in powers of
$\kappa^2$ 
(except for odd $\kappa$'s coming from the CS vertex, see below of). 
At one loop, one would have $\triangle
I_1\sim \kappa^0 \int dx^{11} \triangle L_1$; but there is no
candidate $\triangle L_1$ of dimension $11$, since odd dimension
cannot be achieved  by a purely gravitational $\triangle L_1$,
except at best through a ``gravitational'' $\sim \epsilon\Gamma R
R R R$ or ``form-gravitational'' $\sim \epsilon A R R R R$ CS term
(exemplifying the odd power possibility)\cite{Duff}.However,the
latter term is in fact forbidden by dimension,since it would require 
$3$ further derivatives but is already of {\it even} index order.
The former violates  parity and  hence would represent a (necessarily 
finite) anomaly contribution.
The  two-loop term would be $\triangle L_2\sim\kappa^2\int
d^{11}x \triangle L_2$, so that $\triangle L_2\sim [L]^{-20}$
which can be achieved (to lowest order in external lines) by
$\triangle L_2 \sim \partial^{12} R^4$, where  $\partial^{12}$  
means twelve explicit derivatives spread among the 4 curvatures. 
There are no relevant  $2-$point
$\sim \partial^{16} R^2 $ or  $3-$point $\sim \partial^{14} R^3 $
terms because the $R^2$ can be field-redefined away into the
Einstein action in its leading part (to $h^2$ order, $E_4$ is a
total divergence in any dimension!) while $R^3$ cannot appear by
SUSY. This latter fact was first demonstrated in  D=4 but must
therefore also apply in higher D simply by  the brute force
dimensional reduction argument. So the terms we need are, for
their  4-graviton part, $L^g_4$ of (3) with twelve explicit
derivatives. The companions of $L_4^g$ in $L^{tot}_4$  will simply
appear with the same number of derivatives. It is easy to see that
the additional $\partial^{12}$ can be inserted without spoiling
$SUSY$; indeed they appear as naturally as did multiplication by
$s t u $ in  localizing the $M_4$ to $L_4$: for example,
$\partial^{12}$ might become, in momentum space language,
$(s^6+t^6+u^6)$ or $(stu)^2$. This establishes the  structure of
the $4-$point local counterterm candidate. \\
\indent Before the present construction of the complete counterterm was
achieved, the actual coefficient of its 4-graviton part was
computed \cite{BDetal} by a combination of string-inspired and
unitarity techniques.  Their final result was 
$$ %
{\cal M}_4^{g{\rm twoloop},\ D=11-2\epsilon}\vert_{\rm pole}
= \left( {\kappa \over 2}\right)^6 \times  (stu M_4^{\rm g ~tree}) \times 
$$
$$
  {1\over 48\epsilon\ (4\pi)^{11}} {\pi\over 5791500}
  \Bigl( 438 (s^6+t^6+u^6) - 53 s^2 t^2 u^2 \Bigr) \, ,
$$
where $(stu) M^{\rm g~tree}_4 $ is given in (\ref{pippo}). The
extension of this expression into a counterterm lagrangian for the
rest of the bosonic sector was not presented in \cite{BDetal}, but
is effectively completed here. For detail of \cite{BDetal},
we refer the reader to the review by Bern included in this same volume.
One final comment: nonrenormalizability had always been a
reasonable guess as the fate of D=11 supergravity.
The opposite  guess, however, that some special (M$-$theory related?)
property of this ``maximally maximal'' model might keep it finite
could also have been reasonably
entertained a priori, so this was an issue worth settling.\\
I am deeply  indebted  to my $D=11$ coauthor S. Deser for
many discussions on the subject and suggestions for this
brief review. I am also grateful to Z. Bern, L. Dixon 
and D. Dunbar for stimulating conversation about their
work. This work was supported by  NSF grant PHY-99-73935.


\begin{thebibliography}{99}
\bibitem{CrJuSc} E. Cremmer, B. Julia and J. Scherk, {\it Phys. Lett. B}
{\bf 79} (1978) 409.
\bibitem{BDetal}Z. Bern, L. Dixon, D.C. Dunbar, M. Perelstein and
J.S.  Rozowsky, {\it Nucl. Phys. B}  {\bf 530} (1998) 401.
\bibitem{Green}M. B. Green, M. Gutperle and P. Vanhove, {\it
Phys. Lett. B} {\bf 409} (1997) 177; M. B. Green,  {\it Nucl.
Phys. Proc. Suppl.}  {\bf 68} (1998) 242; M. B. Green, H. Kwon 
and  P. Vanhove, {\it Two loops in Eleven-Dimensions}, hep-th/9910055. 
\bibitem{desergermany} for a review see:  S. Deser, {\it Infinities in
Quantum Gravities}, gr-qc/9911073.
\bibitem{sdds}S. Deser and  D. Seminara, {\it Phys. Rev. Lett.}
{\bf 82} (1999) 2435; and {\it Tree Amplitudes and Two-loop
Countertetrms in $D=11$ Supergravity}, BRX-TH-456 to be submitted.
\bibitem{Sannah} S. Sannan, {\it Phys. Rev. D} {\bf 35} (1987) 1385.
\bibitem{fulling} S. A. Fulling et al., {\it Class. Quant. Grav.} {\bf 9} (1992) 1151.
\bibitem{Schwartz} J. H. Schwarz, {\it Phys. Rept.} {\bf 89} (1982) 223.
\bibitem{DK}S. Deser and J. H. Kay, {\it Phys. Lett. B} {\bf 76}
(1978) 400.
\bibitem{12} J. Plefka, M. Serone and  A. Waldron, {\it
JHEP9811} (1998) 010.
\bibitem{Duff} M. J. Duff and D. Toms, in {``\it Unification of the
fundamental particle interactions II''}, (Plenum Press, New York,
1982).
\end{thebibliography}
\end{document}